\documentclass[conference]{IEEEtran}
\IEEEoverridecommandlockouts
\usepackage{balance}
\usepackage{cite}
\usepackage{float}
\usepackage{amsmath,amssymb,amsfonts}
\usepackage{graphicx}
\usepackage{subcaption}
\usepackage{textcomp}
\usepackage{afterpage}
\usepackage{adjustbox}
\usepackage{stfloats}  
\usepackage{enumitem}
\usepackage{xcolor}
\usepackage{soul}
\sethlcolor{yellow}
\usepackage{tabularx,ragged2e,booktabs}
\usepackage{algorithm}
\usepackage{hyperref}
\usepackage{algorithm,algorithmicx}
\usepackage{algpseudocode}
\usepackage{xfrac}
\usepackage{url}
\usepackage{microtype}
\usepackage{afterpage}
\usepackage{amstext} 
\usepackage{array}   
\newcolumntype{L}{>{$}l<{$}} 

\newcolumntype{C}{>{\raggedright\arraybackslash}X}

\newlength\mylen
\newcolumntype{M}{>{\RaggedRight}m{\mylen}}
\usepackage[switch]{lineno}
\usepackage[T1]{fontenc}
\usepackage{tcolorbox}
\usepackage[scaled=0.85]{beramono}
\usepackage{fancyhdr}
\makeatletter
\algrenewcommand\ALG@beginalgorithmic{\footnotesize}
\algrenewcommand\algorithmiccomment[2][\footnotesize]{{#1\hfill\(\triangleright\) #2}}
\makeatother

\renewcommand{\Comment}[1]{// {\fontfamily{fvm}\selectfont #1}}

\def\BibTeX{{\rm B\kern-.05em{\sc i\kern-.025em b}\kern-.08em
    T\kern-.1667em\lower.7ex\hbox{E}\kern-.125emX}}

\fancypagestyle{alim}{\fancyhf{}\fancyfoot[L]{\footnotesize \textit{The paper has been accepted at the 21st IEEE International Conference on Software Architecture (ICSA) 2024}}}

\begin{document}

\title{Resilient Auto-Scaling of Microservice Architectures with Efficient Resource Management }




\makeatletter
\newcommand{\linebreakand}{%
  \end{@IEEEauthorhalign}
  \hfill\mbox{}\par
  \mbox{}\hfill\begin{@IEEEauthorhalign}
}
\makeatother
 
\author{
  \IEEEauthorblockN{Hussain Ahmad$^{\ast}$}
  \IEEEauthorblockA{\textit{School of Computer and Mathematical Sciences}\\
    \textit{The University of Adelaide, Australia}\\
   hussain.ahmad@adelaide.edu.au}
  \and
  \IEEEauthorblockN{Christoph Treude}
  \IEEEauthorblockA{\textit{School of Computing and Information Systems}\\
    \textit{Singapore Management University, Singapore}\\
    ctreude@smu.edu.sg}
  \linebreakand 
  \IEEEauthorblockN{Markus Wagner}
  \IEEEauthorblockA{\textit{Department of Data Science and AI}\\
  \textit{Monash University, Australia}\\
  markus.wagner@monash.edu}
  \and
  \IEEEauthorblockN{Claudia Szabo}
  \IEEEauthorblockA{\textit{School of Computer and Mathematical Sciences}\\
  \textit{The University of Adelaide, Australia}\\
  claudia.szabo@adelaide.edu.au}
\thanks{$^{\ast}$Corresponding author}
}

\maketitle


\begin{abstract}

Horizontal Pod Auto-scalers (HPAs) are crucial for managing resource allocation in microservice architectures to handle fluctuating workloads. However, traditional HPAs fail to address resource disruptions caused by faults, cyberattacks, maintenance, and other operational challenges. These disruptions result in resource wastage, service unavailability, and HPA performance degradation. To address these challenges, we extend our prior work on Smart HPA and propose Secure\-Smart HPA, which offers resilient and resource-efficient auto-scaling for microservice architectures. Secure\-Smart HPA monitors microservice resource demands, detects disruptions, evaluates resource wastage, and dynamically adjusts scaling decisions to enhance the resilience of auto-scaling operations. Furthermore, Secure\-Smart HPA enables resource sharing among microservices, optimizing scaling efficiency in resource-constrained environments. Experimental evaluation at varying disruption severities, with 25\%, 50\%, and 75\% resource wastage, demonstrates that Secure\-Smart HPA performs effectively across different levels of disruptions. It achieves up to a 57.2\% reduction in CPU overutilization and a 51.1\% increase in resource allocation compared to Smart HPA, highlighting its ability to deliver resilient and efficient auto-scaling operations in volatile and resource-constrained environments.

\end{abstract}


\section{Introduction} \label{section1}

Microservice architectures transform software design by decomposing monolithic systems into modular, independently deployable services, enabling organizations like Netflix, eBay, and Amazon to achieve greater scalability, reliability, and development efficiency \cite{marques2024proactive, jayalath2024microservice}. These services communicate through lightweight interfaces, such as HTTP APIs, ensuring seamless integration across a distributed environment \cite{abdulsatar2024towards}. The deployment of microservices is underpinned by containerization technologies such as Docker \cite{docker} and orchestrated through platforms, such as Kubernetes \cite{kubernetes}, Red Hat OpenShift \cite{RedHat}, and Docker Swarm \cite{DockerSwarm}. Among these, Kubernetes has achieved widespread adoption in both academia and industry due to its advanced features, ecosystem integration, and ability to handle large-scale deployments \cite{rossi2020hierarchical}. A key feature of Kubernetes is the Horizontal Pod Auto-scaler (HPA) \cite{nguyen2020horizontal}, which dynamically adjusts the number of microservice replica pods in response to fluctuating workloads (e.g., Slashdot effect \cite{liu2022coordinating}). HPAs analyze microservice resource usage, workload variations, and Service Level Agreement (SLA) requirements to optimize resource allocation without requiring manual intervention \cite{nguyen2020horizontal}.



Traditional HPAs are constrained by predefined microservice resource limits \cite{nguyen2020horizontal} and prone to HPA architectural vulnerabilities \cite{rossi2020hierarchical}, which compromise their ability to handle dynamic and high-demand workloads \cite{baarzi2021showar}. To address these challenges, we previously proposed Smart HPA \cite{ahmad2024smart}, a resource-efficient HPA that employs a hierarchical architecture that seamlessly integrates centralized and decentralized architectural designs, combining their strengths while minimizing weaknesses. Smart HPA also introduced a resource exchange mechanism that allows microservices to share resources during auto-scaling operations, enabling them to scale beyond predefined resource limits and handle workloads in resource-constrained environments \cite{ahmad2024smart}. However, like traditional HPAs, Smart HPA lacks the capability to detect and adapt to resource disruptions caused by faults, cyberattacks, and other operational challenges. These disruptions often result in resource wastage, reducing the overall capacity available to microservices \cite{zhang2024failure}. For example, a cyberattack can overwhelm a microservice, leading to excessive resource consumption \cite{bremler2024exploiting}, while maintenance activities might temporarily disable its pods. Unfortunately, HPAs are unaware of this wastage and rely on inaccurate resource capacities when making scaling decisions. As a result, HPAs make suboptimal scaling decisions, attempting to provision the desired number of replicas under the assumption that sufficient resources are available. However, due to resource wastage, the desired replicas cannot be fully provisioned, leading to inefficient scaling, degraded performance, and reduced resilience in microservice architectures.


To overcome these limitations, we introduce Secure\-Smart HPA, a novel extension of Smart HPA \cite{ahmad2024smart}, which integrates resilience and resource efficiency into the auto-scaling of microservice architectures. Secure\-Smart HPA enhances the two-layer hierarchical architecture of Smart HPA by introducing a third layer dedicated to identifying and analyzing potential resource disruptions in microservice architectures. This new layer empowers Secure\-Smart HPA to assess the impact of these disruptions and adjust microservice resource capacities, ensuring scaling aligns with actual resource availability for more efficient and resilient operations. The hierarchical architecture combines the advantages of centralized and decentralized designs, enhancing response time while addressing drawbacks like single points of failure. Furthermore, Secure\-Smart HPA employs resilient and resource-efficient heuristics to detect and analyze resource disruptions while reallocating resources among microservices based on real-time microservice resource capacities. By integrating the effects of disruptions and facilitating resource exchange from overprovisioned to underprovisioned microservices, Secure\-Smart HPA maximizes resource utilization. As a result, Secure\-Smart HPA ensures resilience to disruptions in volatile environments while maintaining resource efficiency in constrained environments. To summarize, our contributions are threefold. 
\begin{itemize}[leftmargin=*]
    \item We propose a three-layered hierarchical architecture for Secure\-Smart HPA, integrating centralized and decentralized architectural styles to maximize their strengths and mitigate their limitations, enhancing the management of auto-scaling operations in microservice applications.
    
    \item  We develop resilient and resource-efficient heuristics that detect and analyze resource disruptions while facilitating efficient resource redistribution from overprovisioned to underprovisioned microservices. This enables informed auto-scaling decisions based on available resources in volatile and resource-constrained environments.
    
    \item We evaluate the performance of Secure\-Smart HPA against Smart HPA under varying disruption severity levels: Low, Medium, and High, corresponding to 25\%, 50\%, and 75\% resource wastage, respectively. Our experimental results show that Secure\-Smart HPA performs effectively at all disruption severity levels, achieving up to a 57.2\% reduction in CPU overutilization and a 51.1\% improvement in resource allocation compared to Smart HPA.


\end{itemize}


\section{Background and Related Work} \label{section2}

This section provides an overview of the existing architectures and scaling policies used for HPAs and outlines the distinctive features of Secure\-Smart HPA.

HPA architectures commonly employ the Monitor, Analyze, Plan, Execute, and Knowledge Base (MAPE-K) control framework for microservice auto-scaling operations \cite{marie2020proactive, boyapati2022self, da2021horizontal, ahmad2025towards}. Primarily, HPAs use a centralized MAPE-K control architecture to manage scaling operations in microservice applications \cite{gias2019atom, barna2017delivering, khazaei2017elascale}. This straightforward approach presents challenges such as a single point of failure, higher computational overhead, and limited scalability \cite{rossi2020hierarchical}. A decentralized architecture is proposed to address this, where each microservice has its dedicated auto-scaler \cite{nitto2020autonomic}. For example, Amazon ECS and Kubernetes, prominent container orchestration platforms, employ fully decentralized auto-scalers for microservice scaling operations \cite{nguyen2020horizontal}. Nevertheless, the absence of coordination among auto-scalers results in frequent scaling actions, thereby deteriorating the overall performance of microservices \cite{rossi2020hierarchical}. Recent studies suggest master-worker \cite{rossi2020geo, imdoukh2020machine} and hierarchical \cite{rossi2020self, rossi2020hierarchical} architectures to coordinate decentralized auto-scalers. However, these architectures introduce bottleneck problems and communication overhead during auto-scaling operations \cite{weyns2013patterns}.

HPAs utilize various scaling policies, including threshold-based, fuzzy-based, queuing theory-based, control theory-based, and AI-based approaches, to assess resource demand in response to fluctuating workloads \cite{ahmad2024smart, yu2020microscaler, rossi2020hierarchical}. However, these policies often fail to consider disruptions such as faults or cyberattacks, which can reduce allocated microservice resource capacities, ultimately compromising auto-scaling performance. Also, these policies operate within the pre-allocated resource limits of microservices, making them less effective in resource-constrained scenarios where microservices frequently demand additional resources beyond their initial allocation \cite{ahmad2024smart}.

\noindent\textbf{Distinguishing Features of Secure\-Smart HPA.} The three-layer hierarchical architecture of Secure\-Smart HPA combines centralized and decentralized components to leverage their strengths while addressing limitations. Unlike traditional hybrid architectures, it minimizes communication overhead by engaging centralized components only in resource-constrained scenarios. While existing scaling policies fail to address resource reductions caused by disruptions and remain constrained by predefined resource limits of microservices, Secure\-Smart HPA overcomes these limitations by dynamically updating resource capacities and redistributing resources. This enables it to manage resource reductions caused by disruptions and accommodate resource demands exceeding predefined limits, ensuring resilient and resource-efficient auto-scaling operations. The most closely related work to Secure\-Smart HPA is Smart HPA, a resource-efficient auto-scaler \cite{ahmad2024smart}. Similar to Secure\-Smart HPA, Smart HPA employs a hierarchical architecture to facilitate resource exchange among microservices when resource demands exceed allocated capacities. However, Secure\-Smart HPA enhances the two-layer hierarchical architecture of Smart HPA by introducing a third architectural layer designed to detect and analyze disruptions, update microservice resource capacities, and enable resilient and efficient auto-scaling. This disruption-aware layer equips Secure\-Smart HPA with a unique ability to handle disruptions effectively while optimizing resource utilization, a capability not present in Smart HPA.

\section{Secure\-Smart HPA Architecture} \label{section3}

This section presents the hierarchical architecture and the resilient, resource-efficient heuristics of Secure\-Smart HPA. As depicted in Fig. \ref{Architecture}, the architecture is organized into three core layers: the \textit{Microservice Manager}, \textit{Application Capacity Manager}, and \textit{Application Resource Manager}. Secure\-Smart HPA addresses frequent resource congestion by integrating the MAPE-K framework \cite{weyns2013patterns} into each layer, enabling adaptive auto-scaling of microservice applications. The MAPE-K components are differentiated by subscript notation for each layer: "M" for the Microservice Manager, "C" for the Application Capacity Manager, and "A" for the Application Resource Manager (e.g., Monitor$_M$, Monitor$_C$, and Monitor$_A$). Each layer features a dedicated Knowledge Base that facilitates efficient data sharing among the MAPE components and acts as a repository for scaling decisions, resource metrics, and operational insights. This Knowledge Base ensures situational awareness for DevOps teams responsible for managing Kubernetes-based microservice architectures. In the following, we provide a detailed explanation of each layer, focusing on the functionality of their respective MAPE-K framework components.

\begin{figure*}\vspace{-2mm}%
    \centering
    \includegraphics[width=1.0\linewidth]{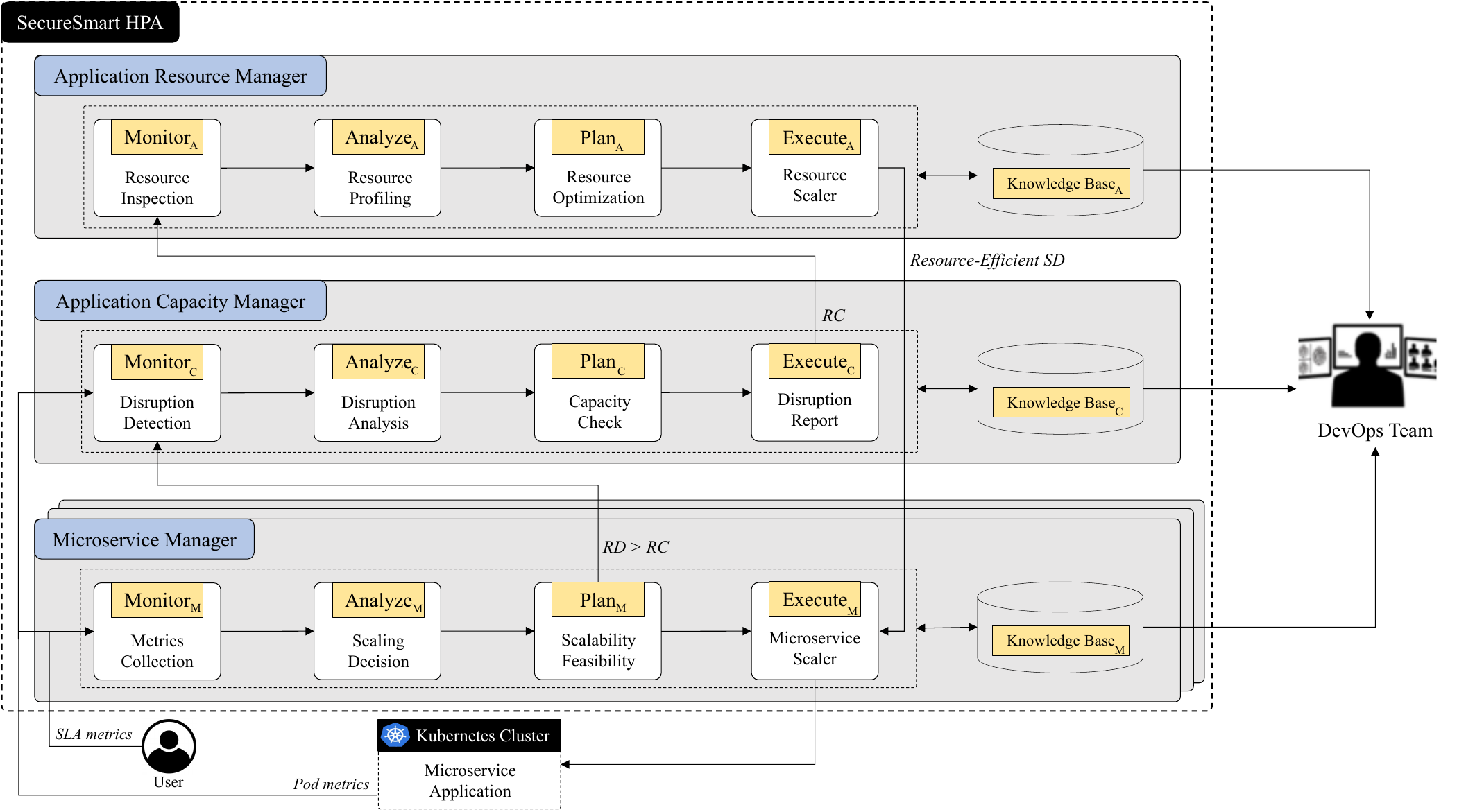}
    \caption{Hierarchical Architecture of Secure\-Smart HPA; RD: Resource Demand, RC: Resource Capacity, SD: Scaling Decision.}
    \label{Architecture}\vspace{1\baselineskip}
\end{figure*}

\begin{algorithm}[t]
\caption{\textsc{Microservice Manager}}\label{alg:MM}
\small
\begin{tabular}{p{8.25cm}}
  \textit{RM:} resource metric value, 
  \textit{RMT:} resource metric threshold value,
  \textit{ResReq:} resource request per replica,
  \textit{CR:} current replica count,
  \textit{DR:} desired replica count,
  \textit{SD:} scaling decision,
  \textit{minR:} minimum replica count,
  \textit{maxR:} maximum replica count,
  \textbf{ACM:} Application Capacity Manager,
  \textit{KB:} knowledge base  \\
  \hline
\end{tabular}
\vspace{2pt}

{\footnotesize$\textbf{Monitor: } RM_i, RMT_i, CR_i, ResReq_i, maxR_i, minR_i $ }
\begin{algorithmic}[1]

\State $DR_i = \text{ceil }(CR_i \times ({RM_i} / {RMT_i}))$ 


\If {$DR_i > CR_i$} 
\State $\textit{SD}_i =  \text{Scale up}$ 
\ElsIf {$DR_i < CR_i \text{ and } DR_i \geq minR_i $} 
\State $\textit{SD}_i =  \text{Scale down}$ 
\Else \ 
\State $\textit{SD}_i =  \text{No Scale}$ 
\EndIf
\If {$DR_i > maxR_i$} 
\State $DR_i, \textit{SD}_i, maxR_i = \textbf{ACM }(DR_i, \textit{SD}_i, CR_i, maxR_i, ResReq_i)$ 
\EndIf
\State $\textbf{Execute } (DR_i, \textit{SD}_i)$
\State $KB \leftarrow DR_i, \textit{SD}_i, CR_i, RM_i, RMT_i, ResReq_i, maxR_i, minR_i$ 
\end{algorithmic}
\end{algorithm}

\subsection{Microservice Manager}

As depicted in Fig. \ref{Architecture}, the Microservice Manager forms the foundational, decentralized layer of the Secure\-Smart HPA hierarchical architecture. Secure\-Smart HPA assigns a dedicated Microservice Manager to each microservice running in a Kubernetes cluster, enabling customized scaling policies and independent adaptation goals. Operating in a decentralized and parallel manner, these managers enhance monitoring accuracy and efficiency, significantly reducing response times compared to a sequential centralized approach. Each Microservice Manager operates using its dedicated MAPE-K framework to gather metrics, analyze data, and execute scaling decisions for its assigned microservice. Algorithm \ref{alg:MM} outlines the detailed functionality of the MAPE-K components for a specific microservice \textit{i}. Below, we detail the role of each MAPE-K component within the Microservice Manager.

\textbf{Monitor$_M$: Metrics Collection.} The Monitor$_M$ component collects key metrics for a microservice, including pod metrics, which indicate the microservice's current status, such as resource utilization ($RM_i$) and the current number of replicas ($CR_i$), as well as SLA metrics that define user requirements, such as resource threshold values ($RMT_i$), minimum ($minR_i$) and maximum replicas ($maxR_i$). These metrics are then forwarded to the Analyze$_M$ component for further processing.

\textbf{Analyze$_M$: Scaling Decision.} The Analyze$_M$ component determines the desired replica count ($DR_i$), detects resource utilization violations (e.g., $RM_i > RMT_i$), and makes scaling decisions ($SD_i$) for its assigned microservice. To implement the Secure\-Smart HPA prototype and ensure a fair comparison with Smart HPA, we employ a static threshold-based scaling policy, similar to Smart HPA \cite{ahmad2024smart}. This policy computes the desired replica count based on the current replica count, resource utilization, and predefined threshold, as shown in line 1 of Algorithm \ref{alg:MM}. A violation is detected when $DR_i$ differs from $CR_i$, indicating that the microservice's resource utilization is outside its defined threshold limits. Once a violation is detected, the Analyze$_M$ component triggers an adaptation process to adjust the replica count, either scaling up or scaling down as necessary. These steps are detailed in Algorithm \ref{alg:MM}, with lines 2–8 outlining violations and corresponding scaling actions.

\textbf{Plan$_M$: Scalability Feasibility.} The Plan$_M$ component determines whether the scaling decision proposed by Analyze$_M$ is feasible within the resource limits of a microservice. If the required resources are within the permissible range (i.e., \textit{$DR_i \leq maxR_i$}), Plan$_M$ forwards the decision to Execute$_M$ for implementation. If the demand exceeds the available capacity (i.e., \textit{$DR_i > maxR_i$}), Plan$_M$ triggers the centralized Application Capacity Manager (ACM) to investigate any disruptions or reductions in microservice resource capacities, as detailed in lines 9-11 of Algorithm \ref{alg:MM}.
 
\textbf{Execute$_M$: Microservice Scaler.} As shown in Line 12 of Algorithm \ref{alg:MM}, the Execute$_M$ component carries out the scaling decision for its assigned microservice by creating the desired replica counts. It is noteworthy that the Microservice Manager stores processed data in the \textbf{Knowledge Base$_M$}, aiding DevOps teams with situational awareness (line 13). Secure\-Smart HPA minimizes communication overhead by operating in a fully decentralized manner when the desired replica count of a microservice remains within its resource capacity, avoiding the need to engage the centralized Application Capacity Manager or Application Resource Manager. However, if the resource demand exceeds capacity, the Application Capacity Manager investigates potential disruptions. Subsequently, the Application Resource Manager redistributes resources among microservices, enabling scaling decisions to be implemented with the available resources to ensure uninterrupted operational continuity.

\begin{algorithm}[t]
\caption{\textsc{Application Capacity Manager}}\label{alg:ACM}
\small
\begin{tabular}{p{8.25cm}}
  \textit{ResLoss:} resource loss,
  \textit{D:} disruption status,
  \textit{pod:} number of available microservice pods in real-time,
  \textit{IRC:} initial resource capacity,
  \textit{CRC:} current resource capacity,
  \textit{KB:} knowledge base  \\
  \hline
\end{tabular}
\vspace{2pt}

{\footnotesize$\textbf{Monitor: } DR_i, \textit{SD}_i, CR_i, maxR_i, ResReq_i, pod_i$}

\begin{algorithmic}[1]
\State $IRC =\sum_{i=1}^M ResReq_i \times maxR_i$ \Comment{Initial Resource Capacity}
\State $CRC =\sum_{i=1}^M ResReq_i \times pod_i$ \Comment{Current Resource Capacity}
\State $ResLoss = IRC-CRC$ 
\If {$ResLoss > 0$} 
\State $D = \text{Disruption Identified}$ 
\State $Severity = ({ResLoss}/{IRC})\times 100$ 
\For {i=1,.., M} \Comment{M = total number of microservices}
\State $maxR_i = pod_i$ \Comment{Update resource capacities}
\EndFor
\Else 
\State $D = \text{No Disruption}$ 
\EndIf
\State $\textbf{Execute } (DR_i, CR_i, SD_i, ResReq_i, maxR_i)$ 
\State $KB \leftarrow D, Severity, maxR_i, IRC, CRC, ResLoss$ 

\end{algorithmic}
\end{algorithm}

\subsection{Application Capacity Manager}

The Application Capacity Manager is built on the MAPE-K framework and acts as the intermediate layer in the Secure\-Smart HPA hierarchical architecture, as shown in Fig. \ref{Architecture}. Its primary role is to monitor and analyze resource capacity reductions in microservice architectures caused by disruptions such as faults, cyberattacks, or operational issues. To mitigate their impact on auto-scaling, the Application Capacity Manager evaluates resource loss and adjusts capacity accordingly. The functionality of the Application Capacity Manager is detailed in Algorithm \ref{alg:ACM}. Below, we provide an explanation of the roles of the MAPE-K components within this layer.

\textbf{Monitor$_C$: Disruption Detection.} The Monitor$_C$ component detects resource reductions caused by disruptions in a microservice application. It first collects the maximum resource capacities ($maxR_i$) for all microservices from the respective Microservice Managers and calculates the Initial Resource Capacities (IRC), as described in line 1 of Algorithm \ref{alg:ACM}. Monitor$_C$ then retrieves the Current Resource Capacities (CRC) by monitoring the total number of available pods for each microservice in real-time on the Kubernetes cluster (line 2). A difference between \textit{IRC} and \textit{CRC} indicates resource loss (line 3), indicating a disruption (lines 4-5) that has reduced the resources available to a microservice application.

\textbf{Analyze$_C$: Disruption Analysis.} Upon detecting a disruption, it is essential to measure its severity, which indicates the proportion of resources lost relative to the total initially allocated capacity. Line 6 of Algorithm \ref{alg:ACM} calculates severity as a percentage by dividing the resource loss by the total initial resource capacity. This analysis is critical for evaluating the impact of disruptions and adjusting resource capacities to enable the Application Resource Manager to make resource-efficient auto-scaling decisions, as detailed in Section \ref{Section:ARM}.

\textbf{Plan$_C$: Capacity Check.} Following the evaluation of disruption severity by Analyze$_C$, Plan$_C$ adjusts the resource capacities of all microservices based on their original resource capacities. This ensures that the actual number of available pods ($pod_i$) in the Kubernetes cluster is accurately reflected in the maximum replicas ($maxR_i$) for each microservice, providing a precise representation of their updated maximum capacity, as outlined in lines 7-9 of Algorithm \ref{alg:ACM}. 

\textbf{Execute$_C$: Disruption Report.} The Execute$_C$ component forwards the updated resource capacities to the Application Resource Manager, which redistributes resources among microservices to maintain auto-scaling operations within the available capacity (line 13 of Algorithm \ref{alg:ACM}). Additionally, the Application Capacity Manager records disruption details, including resource loss and its impact on microservice capacities, in the \textbf{Knowledge Base$_C$} (line 14). This Knowledge Base$_C$ serves as a vital resource for the DevOps team, offering insights into disruptions, their severity, and their effects on resource capacities to support informed decision-making.

\subsection{Application Resource Manager} \label{Section:ARM}

To address the challenge of microservice resource demands exceeding their available capacities, the Application Resource Manager, a centralized component of our hierarchical Secure\-Smart HPA, manages resource redistribution among microservices, ensuring their requirements are met while staying within the overall available resource capacity of the application. To achieve this, we introduce resource-efficient heuristics detailed in Algorithm \ref{algo:ARM}, enabling resource-aware scaling of microservices. The MAPE-K components of the Application Resource Manager are depicted in Fig.~\ref{Architecture} and explained as follows.

\textbf{Monitor$_A$: Resource Inspection.} The Monitor$_A$ component gathers resource information for all microservices within an application running on a Kubernetes cluster. This includes resource demands ($DR_i$), scaling decisions ($SD_i$) made by Microservice Managers, and updated resource capacities ($maxR_i$) provided by the Application Capacity Manager. 


\textbf{Analyze$_A$: Resource Profiling.} To enable efficient resource redistribution, Analyze$_A$ evaluates the monitored data to identify overprovisioned microservices with surplus resources and underprovisioned microservices that require additional capacity. It computes the resource requirements for underprovisioned microservices ($Underprov$), as detailed in lines 1–2 of Algorithm \ref{algo:ARM}, and determines the residual resources available from overprovisioned microservices ($Overprov$), as shown in lines 3–4 of Algorithm \ref{algo:ARM}.

\begin{algorithm}[t]
\caption{\textsc{Application Resource Manager}}\label{algo:ARM}
\small
\begin{tabular}{p{8.25cm}}
  \textit{Overprov:} overprovisioned microservices resource list,
  \textit{Underprov:} underprovisioned microservices resource list,
  \textit{RmaxR:} resource-wise maximum replica count,
  \textit{ResDR:} resource-wise desired replica count,
  \textit{ResSD:} resource-wise scaling decision \\
  \hline
\end{tabular}
\vspace{2pt}

{\footnotesize$\textbf{Monitor: } DR_i, SD_i, CR_i, maxR_i, ResReq_i $ }

\begin{algorithmic}[1]
\If {$DR_i > maxR_i$} 
\State $Underprov.\text{append}((DR_i-maxR_i)\times ResReq_i)$ 
\Else
\State  $Overprov.\text{append}((maxR_i-DR_i)\times ResReq_i)$ 
\EndIf
\State $Underprov \gets \text{Dsort}(\textit{Underprov})$ \Comment{Most to least underprovision}
\State $Overprov \gets \text{Dsort}(\textit{Overprov})$ \Comment{Most to least overprovision}
\For{$i \in Overprov$} 
\State Extract needful resource from $i$-th service 
\State $RmaxR_i =$ Update $maxR_i$ 
\EndFor
\For{$i \in Underprov$} 
\State Add required resource to $i$-th service
\State $RmaxR_i =$ Update $maxR_i$ 
\EndFor
\For{$i =1 \text{ to } Underprov$} 
\If{$RmaxR_i >= DR_i$}
\State $ResSD_i = SD_i$ , $ResDR_i = DR_i$ 
\ElsIf{$RmaxR_i \in [maxR_i, DR_i]$} 
\State $ResSD_i =$ Scale up , $ResDR_i = RmaxR_i$ 
\Else
\State $ResSD_i =$ No Scale , $ResDR_i = CR_i$ 
\EndIf
\EndFor
\State $\textbf{Execute } (ResDR_i, ResSD_i, RmaxR_i)$ 
\State $KB \leftarrow ResDR_i, ResSD_i, RmaxR_i$ 

\end{algorithmic}
\end{algorithm}

\textbf{Plan$_A$: Resource Optimization.} The Plan$_A$ component manages the redistribution of resources by transferring residual capacity from overprovisioned to underprovisioned microservices, dynamically adjusting their resource capacities to meet their demands. As outlined in lines 6–15 of Algorithm \ref{algo:ARM}, this process begins by prioritizing microservices with the highest resource needs, focusing on the most underprovisioned first. Therefore, underprovisioned microservices ($Underprov$) are sorted in descending order of their resource demands (line 6), while overprovisioned microservices ($Overprov$) are sorted in descending order of their residual resources to ensure surplus is extracted from those with the greatest excess (line 7). Resources are then redistributed iteratively, drawing from the most overprovisioned microservices and allocating to the most underprovisioned first (lines 8–15). This process continues until either all underprovisioned microservices' demands are met or the overprovisioned microservices have no residual resources left. Throughout this process, the Plan$_A$ component continuously updates the capacities of overprovisioned and underprovisioned microservices, reflecting reductions and increases in their capacities as needed, as outlined in lines 10 and 14 of Algorithm \ref{algo:ARM}. If residual resources are insufficient to fully meet the demands of underprovisioned microservices, Plan$_A$ allocates the available surplus from overprovisioned microservices to maximize support for the most underprovisioned ones. If no residual resources remain, redistribution does not occur, and the existing allocation is preserved.

\textbf{Execute$_A$: Resource Scaler.} After the Plan$_A$ component updates resource capacities, Execute$_A$ adjusts resource-wise desired replica counts and scaling decisions for underprovisioned microservices, as outlined in lines 16–24 of Algorithm \ref{algo:ARM}. These resource-wise scaling decisions ($ResSD_i$), desired replicas ($ResDR_i$), and resource capacities ($RmaxR_i$) are communicated to the respective Execute components of Microservice Managers for implementation (Fig. \ref{Architecture}). The Application Resource Manager stores values of $ResDR_i$, $ResSD_i$, and $RmaxR_i$ in the \textbf{Knowledge Base$_A$}, providing DevOps Team real-time insights to facilitate effective monitoring and decision-making.

It is worth highlighting that the proposed resilient and resource-efficient heuristics are designed to integrate seamlessly with other scaling policies (e.g., queuing theory) and performance metrics (e.g., CPU usage and response time). This flexibility allows researchers and practitioners to tailor the heuristics to specific application needs, enabling customizable, resilient, and efficient auto-scaling.

\section{Experimental Evaluation} \label{section4}

We evaluate Secure\-Smart HPA through two key research questions: \textit{RQ1: What is the performance of Secure\-Smart HPA under varying levels of disruption severity? RQ2: What performance improvements does Secure\-Smart HPA achieve over Smart HPA under varying levels of disruption severity?} To answer these questions, we first outline the experimental setup designed to assess Secure\-Smart HPA, followed by an in-depth analysis of the evaluation results.

\subsection{Experimental Setup} \label{section4.1}

This section provides a detailed overview of the experimental setup, covering the environment configuration, benchmark microservice application, load testing parameters, evaluation metrics, experimental scenarios, and the disruption injection process.

\subsubsection{Experiment Environment}

To evaluate the performance of Secure\-Smart HPA, we utilize Amazon Web Services (AWS) \cite{amazon} to host a benchmark microservice application on 15 virtual machines (VMs), each configured as Amazon EC2 \textit{t3.medium} instances with 2-core Intel Xeon Platinum 8000 processors, 3.1 GHz CPU, 4 GB RAM, 5 Gbps network bandwidth, and 5 GiB disk storage. These VMs, running the EKS-optimized Amazon Linux 2 operating system (AL2\_\texttimes{}86\_64), are orchestrated using Amazon Elastic Kubernetes Service (EKS) \cite{kube} with Kubernetes version 1.31. The EKS cluster employs default AWS Virtual Private Cloud (VPC) settings, an IPv4 cluster family, API endpoints accessible via public and private networks, and critical add-ons such as kube-proxy, CoreDNS, and VPC CNI. Secure\-Smart HPA is deployed locally on a machine equipped with a 2.60 GHz Intel Core i7 processor and 16 GB RAM, communicating with the AWS-hosted application via the AWS command-line interface. For implementing distributed microservice managers of Secure\-Smart HPA, we use \textit{Python’s multiprocessing module} \cite{aziz2021python}, enabling each auto-scaling operation to run as an independent process with its own memory space. This design avoids the limitations of \textit{Python’s Global Interpreter Lock} found in the threading module and enhances the efficiency of auto-scaling actions \cite{aziz2021python}.

\begin{figure}[t]\vspace{-2mm}%
  \hspace*{0.3cm}
  \includegraphics[trim=90 170 190 130,clip, scale = 0.375]{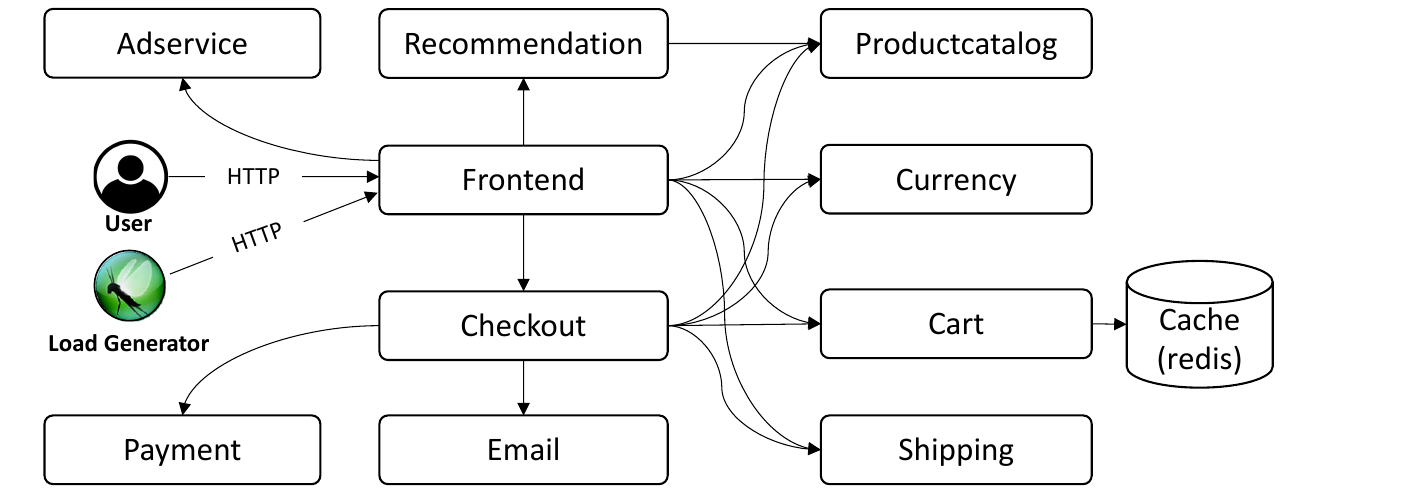}
  \caption{Online Boutique Architecture \cite{boutique}.}
  \label{fig:BenchmarkApp} 
  \vspace{-1.0\baselineskip}
\end{figure}

{\footnotesize
\begin{table}[b]
\vspace{-1.0\baselineskip}  
\caption{\centering Evaluation Metrics. \newline {\small{milliCPU: 1/1000th of a core; CPU Demand: required resources; CPU Capacity: total available resources.}}}
\label{table:EvaluationMetrics}
\begin{tabularx}{88mm}{>{\hsize=0.3\hsize}X>{\hsize=0.7\hsize}X}
\toprule

\textbf{Evaluation Metric} & \multicolumn{1}{c}{\textbf{Description}} \\ \midrule

Supply CPU \newline(milliCPU) & CPU resources allocated to the current replicas of microservices in an application. \\ 

CPU Overutilization (percent usage) & CPU usage exceeding a predefined threshold (CPU Surpass - CPU Threshold). \\ 

CPU Underprovision \newline(milliCPU) & Unavailable CPU resources required by an application (CPU demand - CPU capacity). \\

CPU Overprovision \newline(milliCPU) & Unused allocated CPU resources of a microservice application (CPU capacity - CPU demand). \\ 

\bottomrule

\end{tabularx}
\vspace{-1.5\baselineskip}  
\end{table}
}

\subsubsection{Benchmark Microservice Application} \label{benchmarkk}

Secure\-Smart HPA is easy to use and can work with any microservice application running on Kubernetes, making it flexible for diverse microservice applications. However, to evaluate the resilience and resource efficiency of Secure\-Smart HPA, we employ the Online Boutique application \cite{boutique}, a widely recognized microservice benchmark that aligns with established benchmark selection criteria \cite{aderaldo2017benchmark}. This benchmark is extensively used in academic and industry research to advance knowledge and innovation in microservice architectures \cite{santos2023gym, choi2021phpa, karn2022automated}. The Online Boutique is an e-commerce platform that enables users to browse items, manage shopping carts, and complete purchases (Fig. \ref{fig:BenchmarkApp}). The application is composed of 11 interconnected microservices, each written in different programming languages, providing a complex and realistic test environment. For this study, we use the default configuration of the Online Boutique, maintaining CPU settings from Smart HPA \cite{ahmad2024smart} for a fair comparison with Secure\-Smart HPA. Most services are allocated CPU requests and limits of 100m/200m, with exceptions for \textit{cart} and \textit{adservice} at 200m/300m, and \textit{redis} at 70m/125m. To streamline the deployment and minimize network overhead, Docker images were preloaded onto each virtual machine.


\subsubsection{Application Load Testing} \label{loadtestingg}

The application package of Online Boutique provides a load-testing script to evaluate its scalability by simulating user interactions such as browsing, exploring products, and completing transactions \cite{boutique}. The load test is executed using the Locust load testing tool \cite{locust}, which runs on the same machine as Secure\-Smart HPA. As presented in Fig. \ref{fig:LoadTest}, Locust sends HTTP requests to the benchmark application hosted on AWS EKS, simulating users (Fig.\ref{fig:Simulated_Users}) and corresponding workload patterns (Fig. \ref{fig:Simulated_Workload}). The load test runs for 15 minutes, beginning with a 5-minute ramp-up phase where the number of users increases from zero to 600 at a rate of one user every two seconds (2-second spawn rate). This phase examines the ability of Secure\-Smart HPA to handle increasing workloads. It is followed by a 10-minute period of sustained high load, with all 600 users actively generating traffic to simulate a resource-constrained scenario. Since real-world load profiles typically involve periods of increasing and sustained high loads, this load profile, combining a ramp-up and sustained phase, reflects common traffic patterns, ensuring realistic and effective scalability testing.

\subsubsection{Evaluation Metrics}

To evaluate the resilience and resource efficiency of Secure\-Smart HPA, we use \textit{CPU utilization} as the scaling metric to assess resource demand, capacity, surpluses, and shortages in microservices, providing insights into auto-scaling performance. Table~\ref{table:EvaluationMetrics} presents the evaluation metrics used to evaluate Secure\-Smart HPA. For example, CPU Underprovision indicates resource shortages, while CPU Overprovision quantifies excess capacity. Additionally, we align our scaling metric with Smart HPA \cite{ahmad2024smart} to ensure a fair comparison between Secure\-Smart HPA and Smart HPA.



\begin{figure}[t]
  \centering\vspace{-2mm}
  \begin{minipage}[b]{0.49\linewidth}
    \includegraphics[height=0.8in,width=1.65in]{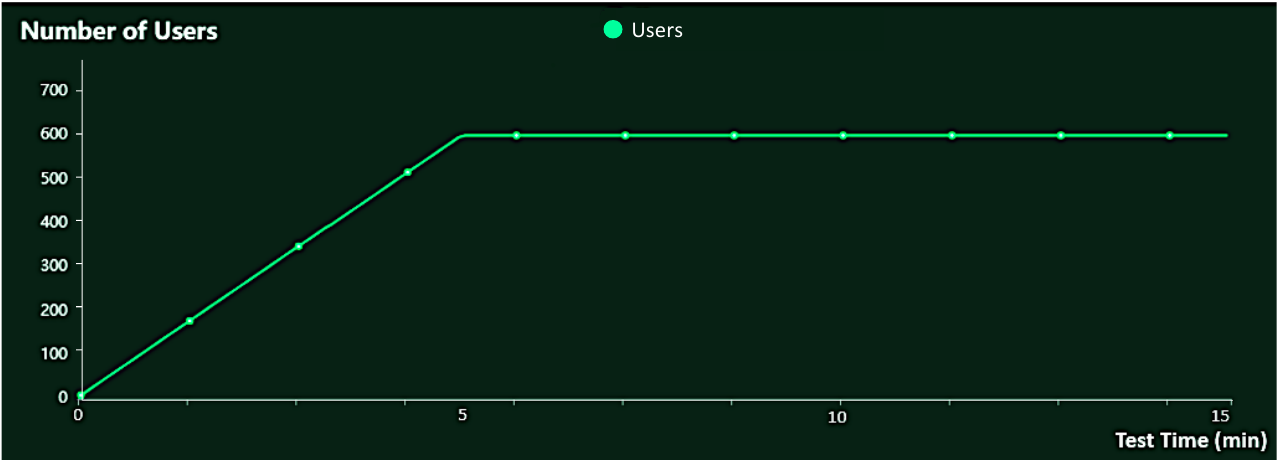}
    \subcaption{Simulated Users.}
    \label{fig:Simulated_Users}
  \end{minipage}
  \hfill
  \begin{minipage}[b]{0.49\linewidth}
    \includegraphics[height=0.8in,width=1.65in]{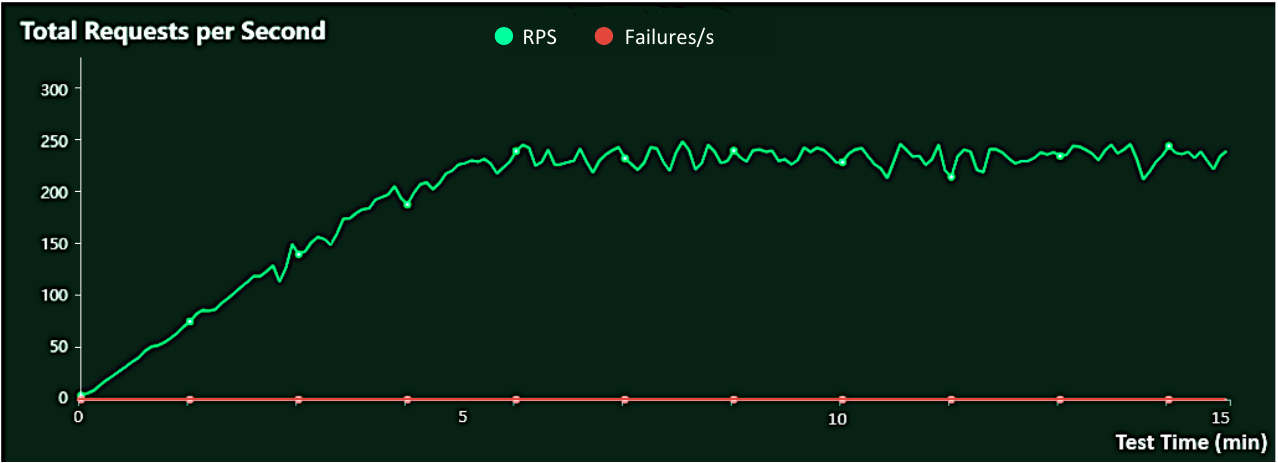}
    \subcaption{Simulated Workload.}
    \label{fig:Simulated_Workload}
  \end{minipage}\vspace{-1.0mm}
  \caption{Load Test for Benchmark Application.}
  \label{fig:LoadTest}
    \vspace{-3mm}
\end{figure}

\subsubsection{Experimental Scenarios}


To evaluate the performance of Secure\-Smart HPA, we design experimental scenarios representing different levels of disruption severity: Low, Medium, and High. These scenarios correspond to resource wastage levels of 25\%, 50\%, and 75\%, respectively, simulating varying degrees of impact on microservice resources. Table~\ref{table:experimental_scenarios} presents the designed experimental scenarios, along with the corresponding available resource levels for the benchmark microservice application used in evaluating Secure\-Smart HPA. We conduct the experiments with the benchmark microservice application configured to have a maximum of 5 replicas per service and a 50\% CPU utilization threshold (referred to as 5R-50\%). According to the resource settings of the benchmark application described in Section \ref{benchmarkk}, this configuration results in a total CPU resource capacity of 6350 mCPU for the 11 microservices. This setup represents the most favorable scenario for Smart HPA when compared to the Kubernetes baseline HPA \cite{ahmad2024smart}. By selecting this optimal configuration, we ensure a fair comparison that highlights the performance improvements introduced by Secure\-Smart HPA.


\begin{table}[t]
    \centering
    \caption{\centering Experimental Scenarios. \newline {\small Total Resource Capacity: 6350 mCPU.}}
    \begin{tabular}{lll}
        \toprule
        \textbf{Disruption Severity} & \textbf{Resource Wastage} & \textbf{Remaining Resource} \\
        \midrule
        Low & 25\% & 4762.5 mCPU \\
        Medium & 50\% & 3175 mCPU \\
        High & 75\% & 1587.5 mCPU \\
        \bottomrule
    \end{tabular}
    \label{table:experimental_scenarios}
\vspace{-1.5\baselineskip}
\end{table}

\subsubsection{Disruption Injection}

Disruptions in a microservice application result in resource wastage during auto-scaling operations. To inject such disruptions, we randomly delete microservice pods to achieve the desired resource wastage levels specified in the experimental scenarios. The random selection of microservices and the number of replicas to delete are determined using Python’s \textit{random} function, which selects a microservice from the 11 services in the benchmark application and decides the number of replicas to remove. For instance, in a Medium-severity disruption, we delete microservice pods consuming a total of 3175 mCPU (50\% of the total resources), ensuring that only 50\% of the resources remain available for the application. A critical consideration is when to introduce a disruption during the 15-minute load test described in Section \ref{loadtestingg}. Disruptions are more likely to occur when an application is fully utilized, meaning all services are actively handling peak loads. According to the load test profile, the number of users increases from 0 to 600 in the first 5 minutes, followed by 600 concurrent users sending HTTP requests for the next 10 minutes. Therefore, we inject the disruption at the 5.5-minute mark, just after the benchmark application reaches full load. This timing ensures that the application is busy and provides enough observation time (9.5 minutes) to analyze the impact of disruption on the evaluation metrics.

\begin{figure*}[t]\vspace{-2mm}
    \begin{subfigure}[b]{0.245\textwidth} 
        \includegraphics[ trim=.3cm .7cm 0.5cm .68cm, clip,width=1.\textwidth]{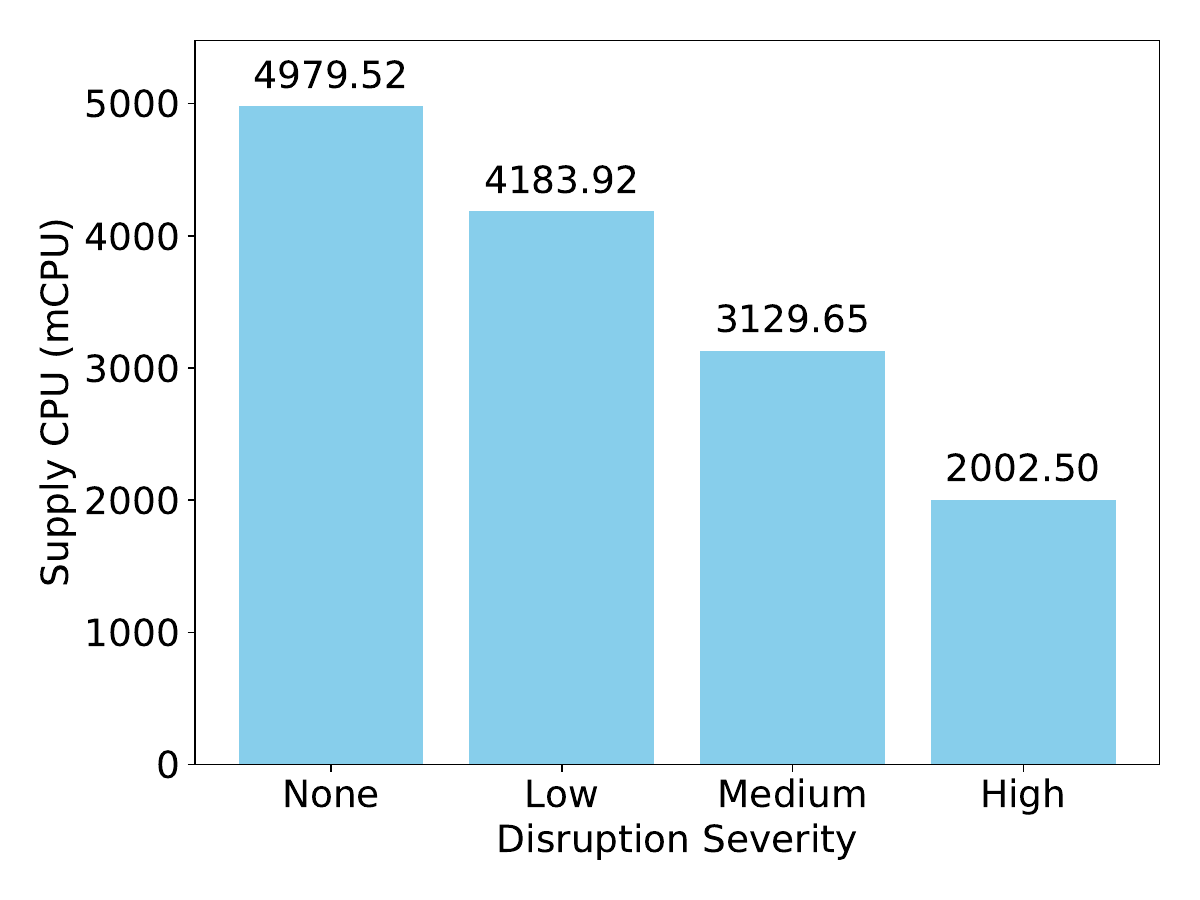} 
        \caption{Supply CPU}
        \label{fig:fig1}
    \end{subfigure}
    \begin{subfigure}[b]{0.245\textwidth}
        \includegraphics[trim=0.3cm .7cm 0.5cm .68cm, clip,width=1.\textwidth]{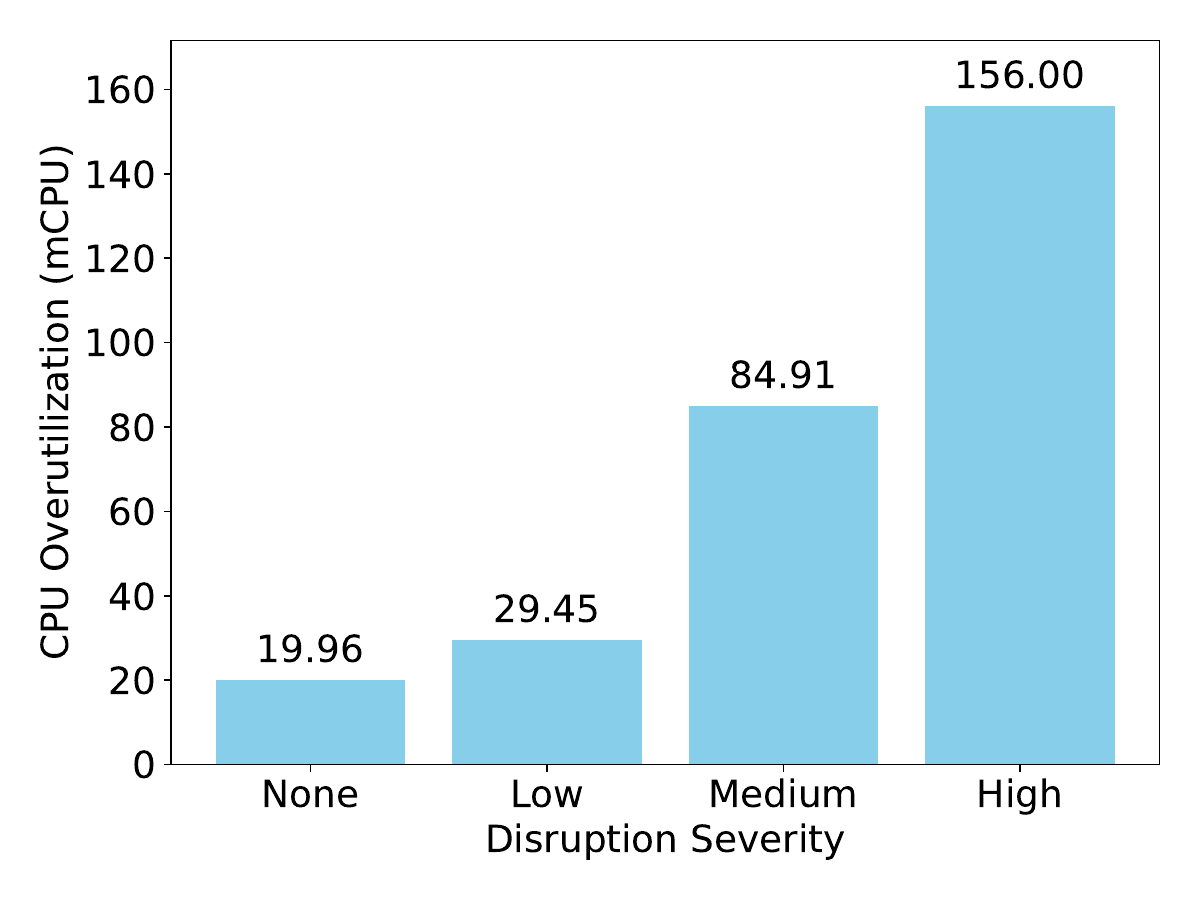} 
        \caption{CPU Overutilization}
        \label{fig:fig2}
    \end{subfigure}
    \begin{subfigure}[b]{0.245\textwidth}
        \includegraphics[trim=.3cm .7cm 0.5cm .68cm, clip,width=1.\textwidth]{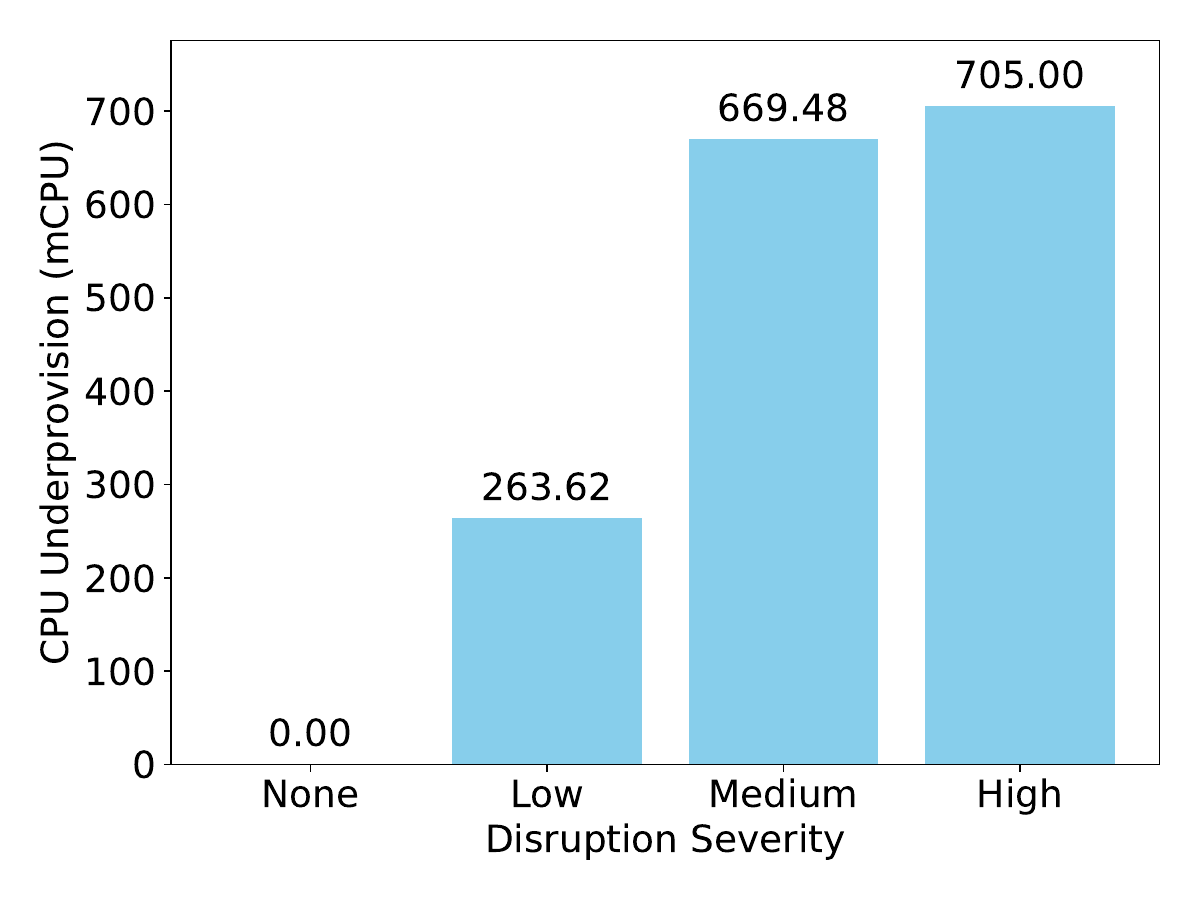} 
        \caption{CPU Underprovision}
        \label{fig:fig3}
    \end{subfigure}
    \begin{subfigure}[b]{0.245\textwidth}
        \includegraphics[trim=.3cm .7cm 0.5cm .68cm, clip,width=1.\textwidth]{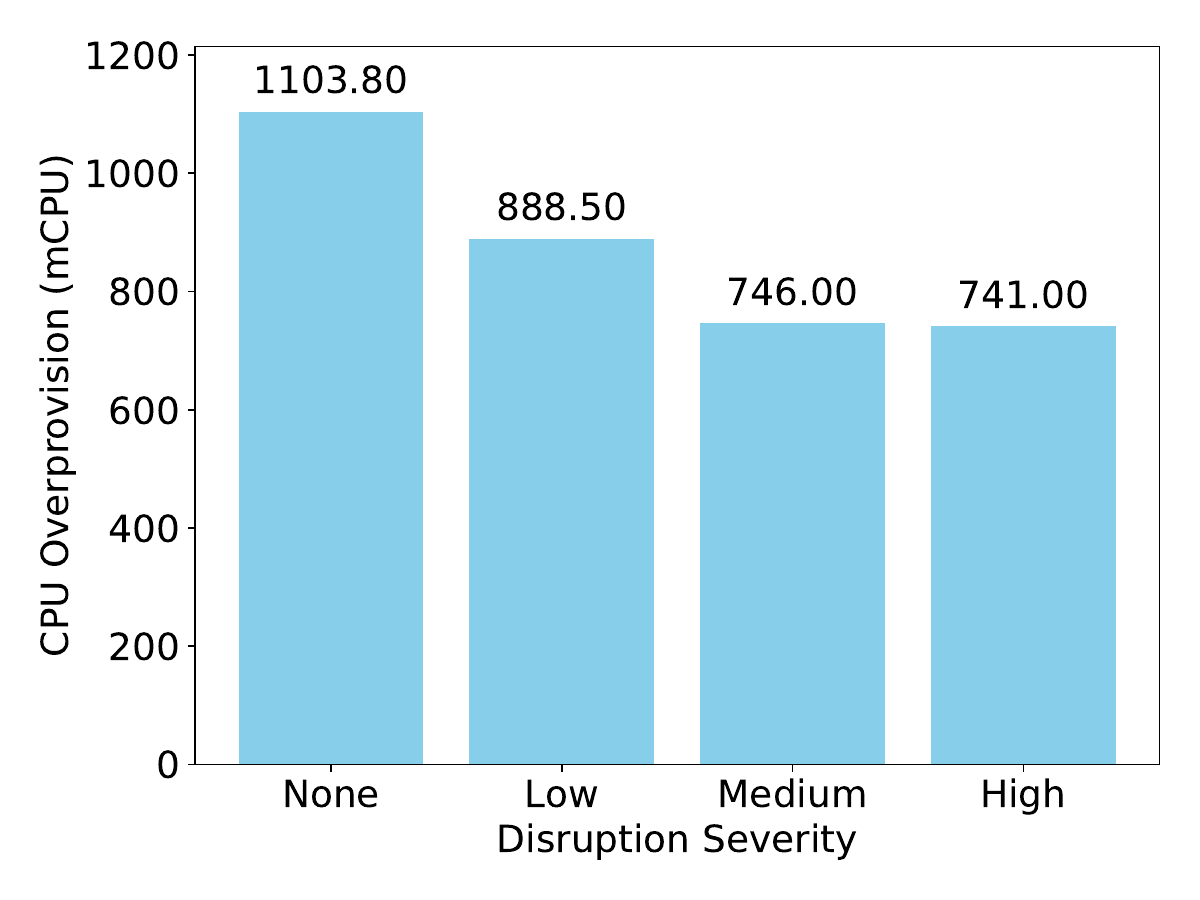} 
        \caption{CPU Overprovision}
        \label{fig:fig3}
    \end{subfigure}
    
    \caption{Performance of Secure\-Smart HPA Across Varying Disruption Severity Levels.}
    \label{fig:results1}
    \vspace{2mm}
\end{figure*}

\subsection{Results and Discussion} \label{section4.2}

This section presents the results of our experimental evaluation of Secure\-Smart HPA. To assess its performance, we conduct extensive load testing on the benchmark application managed by Secure\-Smart HPA across three experimental scenarios, each characterized by a different level of disruption severity: Low (25\% resource wastage), Medium (50\% resource wastage), and High (75\% resource wastage). To ensure the reliability of our findings, each load test is repeated 10 times per scenario, and the average results are computed.


\subsubsection{Secure\-Smart HPA Performance Across Varying Disruption Severity Levels} \label{section4.2.1}

This section addresses our first research question: \textit{RQ1: What is the performance of Secure\-Smart HPA under varying levels of disruption severity?} Fig.~\ref{fig:results1} shows the evaluation metrics for the benchmark application deployed on AWS EKS with Secure\-Smart HPA, under scenarios with no disruptions and with Low, Medium, and High severity disruptions during load testing. The \textbf{Supply CPU} metric, which represents allocated resources, decreases as disruption severity increases. In case of \textbf{no severity}, the allocated CPU is 4979.52 mCPU. Under \textbf{Low severity}, it decreases to 4183.92 mCPU, a 15.9\% reduction. In the \textbf{Medium severity} scenario, it further drops to 3129.65 mCPU, representing a 25.2\% decrease from Low severity. At \textbf{High severity}, the available CPU falls to 2002.50 mCPU, marking a 36.0\% decrease from the Medium level. These results highlight that while Secure\-Smart HPA strives to maintain application operations by allocating CPU resources during disruptions, the Supply CPU significantly declines as severity intensifies, leaving microservices critically constrained at higher disruption levels. This decline in Supply CPU is further reflected in the \textbf{CPU Overutilization} metric, which measures the percentage of CPU usage exceeding the utilization threshold of 50\%. At \textbf{no severity}, CPU Overutilization is 19.96 mCPU, indicating minimal strain. It increases to 29.45 mCPU under \textbf{Low severity}, a 47.5\% rise. At \textbf{Medium severity}, it reaches 84.91 mCPU, a 188.3\% increase from Low. Under \textbf{High severity}, overutilization peaks at 156.00 mCPU, reflecting an 83.7\% increase from Medium. The relatively small increase from no disruption to Low disruption indicates efficient resource management with residual resources, while the sharp rise at Medium and High disruption levels highlights escalating contention as resources become increasingly scarce.

Disruptions significantly impact both CPU underprovision and overprovision. The \textbf{CPU Underprovision} metric represents the CPU resources that microservices demand but are not available. Under \textbf{no severity}, the underprovision is 0.00 mCPU, indicating that Secure\-Smart HPA fully meets demand. It rises to 263.62 mCPU under \textbf{Low severity}. At \textbf{Medium severity}, it reaches 669.48 mCPU, a 154\% increase from Low. At \textbf{High severity}, it climbs to 705.00 mCPU, representing a 5.3\% increase from Medium. This trend highlights the growing challenges Secure\-Smart HPA faces in resource allocation as disruption severity escalates. Higher resource wastage reduces the resources available for redistribution, making it increasingly difficult for Secure\-Smart HPA to manage workloads effectively. The \textbf{CPU Overprovision} metric, which measures the residual CPU resources that remain unused, shows a consistent decrease as disruptions intensify. At \textbf{no severity}, the CPU Overprovision is 1103.80 mCPU. It decreases to 888.50 mCPU at \textbf{Low severity}, reflecting a 19.5\% reduction. At \textbf{Medium severity}, it drops to 746.00 mCPU, a 16.0\% decrease from Low. At \textbf{High severity}, it stabilizes at 741.00 mCPU, showing a marginal 0.67\% decrease from Medium. This trend indicates that Secure\-Smart HPA minimizes residual resources to enhance auto-scaling efficiency in response to increasing resource wastage. However, the minimal decrease from Medium to High severity suggests that nearly all residual resources have already been allocated, leaving little to no residual capacity. The final 741.00 mCPU represents the residual resource just before the 5.5-minute mark of the load test, at which point the disruption is introduced.

\begin{tcolorbox}[left=0pt, top=0pt, right=0pt, bottom=0pt, boxrule=0.75pt]

Secure\-Smart HPA exhibits adaptive, resilient, and resource-efficient behavior by effectively managing resource constraints at different disruption severity levels, ensuring the continuity of auto-scaling operations with available resources.

\end{tcolorbox}


\begin{figure}\vspace{-2mm}%
\includegraphics[trim=1.5cm 0.2cm 2cm 1.5cm, clip,width=1.0\linewidth]{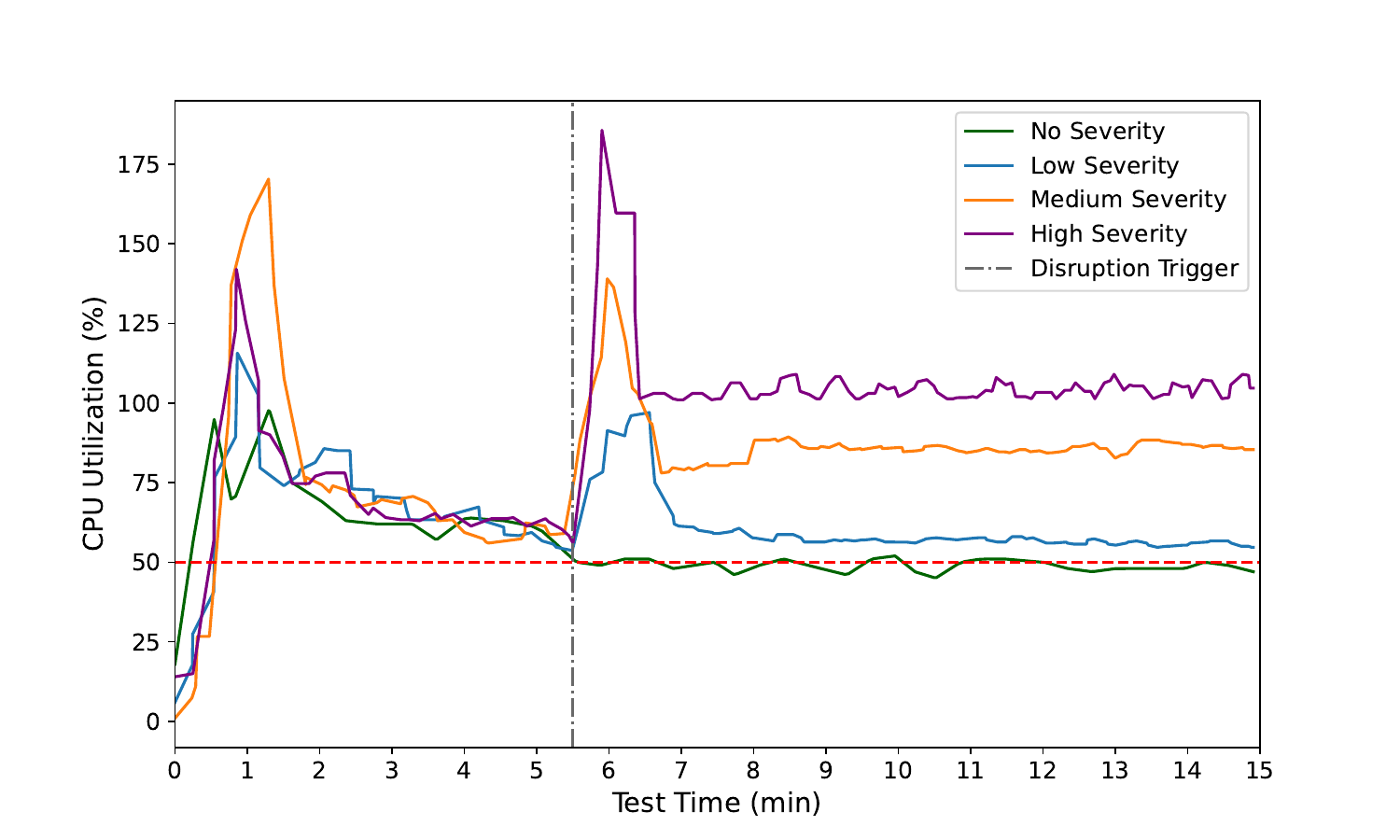}
\caption{\centering Secure\-Smart HPA Adaptability during Load Test.}
\label{fig:results2}
\vspace{-1.5\baselineskip}
\end{figure}

\noindent \textbf{Adaptability of Secure\-Smart HPA Across Varying Disruption Severity Levels.} Fig. \ref{fig:results2} illustrates the CPU Utilization of the benchmark microservice application under Secure\-Smart HPA during a 15-minute load test, with disruptions introduced at the 5.5-minute mark to simulate Low (25\% resource wastage), Medium (50\% resource wastage), and High (75\% resource wastage) severity levels. The red dashed line represents the CPU threshold of 50\%. In the \textbf{no severity} scenario, where no disruptions are introduced, CPU utilization remains stable and aligned with the 50\% threshold throughout the test.  In the \textbf{Low severity} scenario, CPU utilization temporarily spikes after the 5.5-minute disruption but quickly stabilizes near the 50\% threshold,  settling around 60\%. This reflects the efficiency of Secure\-Smart HPA's Application Capacity Manager, which promptly detects and evaluates the disruption, adjusting resource capacities across all microservices. The Application Resource Manager then redistributes the remaining resources, reallocating them from overprovisioned to underprovisioned microservices to sustain performance. This adaptive response ensures the application operates effectively within the reduced resource capacity, even with a 25\% resource loss. In the \textbf{Medium severity} scenario, the spike in CPU utilization following the disruption is more pronounced,  stabilizing at a higher level of around 90\% compared to the Low severity scenario. CPU utilization consistently exceeds the 50\% threshold, indicating heightened resource contention caused by the 50\% reduction in available resources. While Secure\-Smart HPA effectively mitigates excessive overutilization, it begins to show signs of strain as it manages a heavier workload with significantly fewer resources, highlighting the increased challenges in maintaining performance under these conditions. In the \textbf{High severity} scenario, the CPU utilization shows a sharp peak after the disruption, reaching above 175\%, before stabilizing at around 110\%. The benchmark application operates far above the 50\% CPU threshold, indicating that the remaining 25\% of resources are insufficient to meet demand fully. However, despite this significant resource loss, Secure\-Smart HPA prevents complete service failure, demonstrating its resilience and ability to maintain functionality under severe resource constraints.

\begin{tcolorbox}[left=0pt, top=0pt, right=0pt, bottom=0pt, boxrule=0.75pt]

Secure\-Smart HPA stabilizes CPU utilization near the threshold following disruptions through its resilient resource exchange mechanism. However, as disruption severity escalates, the deviation from the threshold becomes more pronounced, reflecting the increasing complexity of resource management under significant resource losses.
\end{tcolorbox}

\begin{figure}[t]\vspace{-1mm}%
    \centering
    \begin{subfigure}[b]{0.235\textwidth}
        \centering
        \includegraphics[trim={0.7cm 0.65cm 0.65cm 0.65cm},clip,width=\linewidth]{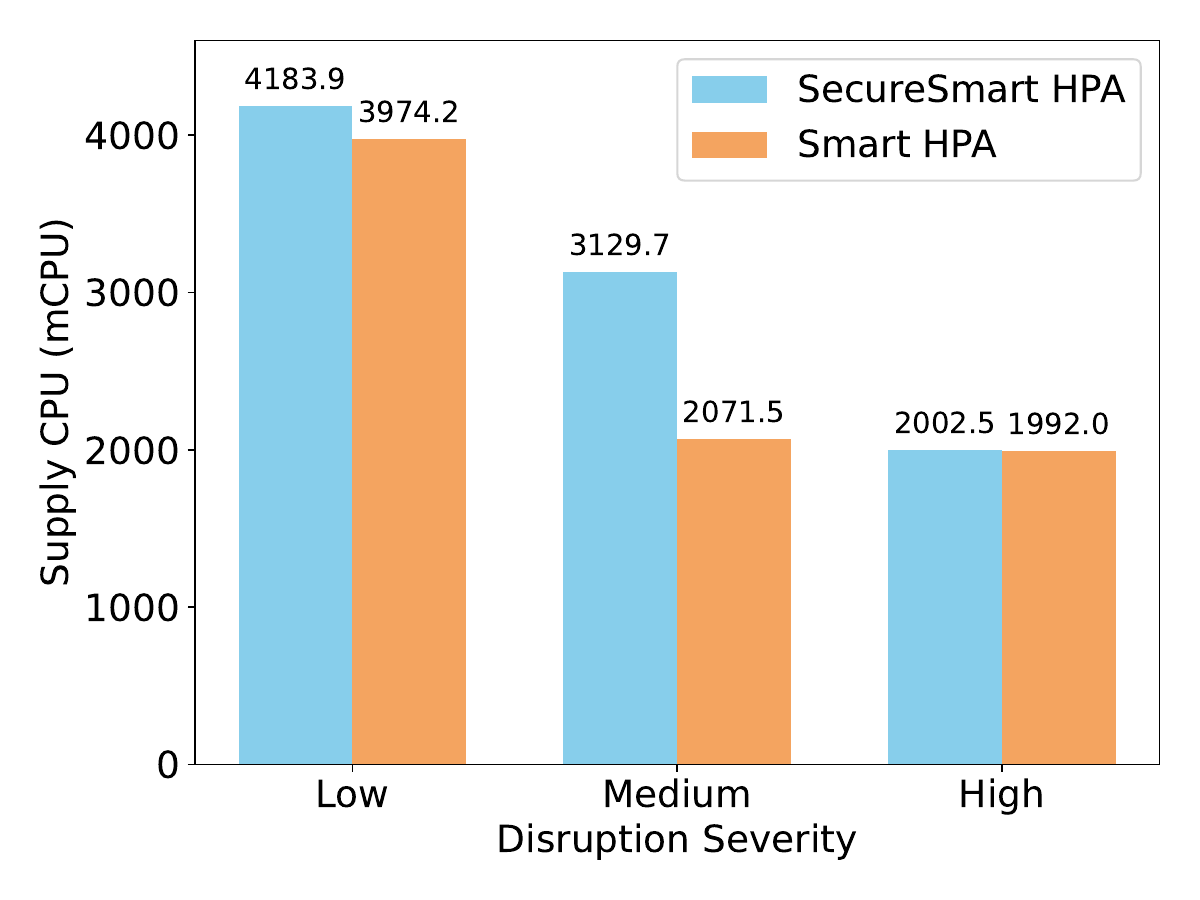}
        \caption{Supply CPU}
    \end{subfigure}
    \hspace{0.01cm}
    \begin{subfigure}[b]{0.235\textwidth}
        \centering
        \includegraphics[trim={0.7cm 0.65cm 0.65cm 0.65cm},clip,width=\linewidth]{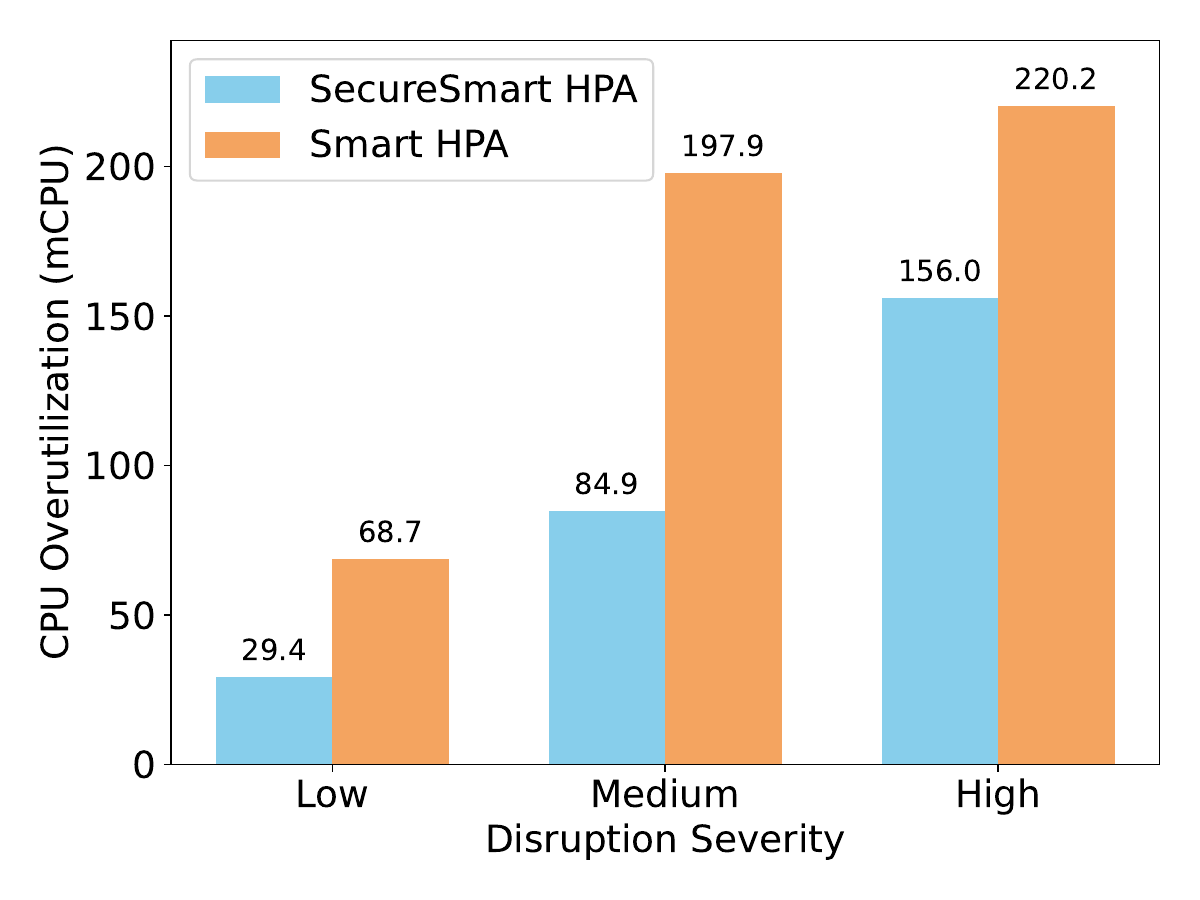}
        \caption{CPU Overutilization}
    \end{subfigure}
    \caption{Comparison between Secure\-Smart HPA and Smart HPA.}
    \vspace{-1.5\baselineskip}
    \label{fig:compSSvS}
\end{figure}
\subsubsection{Comparison between Secure\-Smart HPA and Smart HPA} \label{section4.2.2}

This section addresses our second research question: \textit{RQ2: What performance improvements does Secure\-Smart HPA achieve over Smart HPA under varying levels of disruption severity?} Fig. \ref{fig:compSSvS} compares Secure\-Smart HPA and Smart HPA across Low, Medium, and High severity levels, highlighting an enhanced performance of Secure\-Smart HPA in managing resource allocation after disruptions. At \textbf{Low severity} (25\% resource wastage), Secure\-Smart HPA supplies 4183.92 mCPU, outperforming Smart HPA’s 3974.16 mCPU by 5\%. Additionally, it reduces CPU Overutilization to 29.45 mCPU, compared to 68.74 mCPU for Smart HPA, achieving a 57.2\% improvement. These results demonstrate Secure\-Smart HPA's efficiency in managing minor disruptions by promptly detecting and analyzing them, adjusting microservice resource capacities to account for resource losses, and redistributing resources effectively. For \textbf{Medium severity} (50\% resource wastage), Secure\-Smart HPA provides 3129.65 mCPU, outperforming Smart HPA, which supplies only 2071.50 mCPU, a 51.1\% increase in Supply CPU. Additionally, Secure\-Smart HPA limits CPU Overutilization to 84.91 mCPU, compared to 197.93 mCPU for Smart HPA, delivering a 57.1\% reduction in CPU Overutilization. This highlights the capability of Secure\-Smart HPA to efficiently manage resources as disruptions become more severe. In contrast, the lack of disruption monitoring in Smart HPA leads to incorrect identification of overprovisioned and underprovisioned microservices, resulting in inefficient resource exchange. This inefficiency causes an increase in CPU Overutilization while also limiting the effective utilization of the available CPU supply. At \textbf{High severity} (75\% resource wastage), Secure\-Smart HPA maintains a supply of 2002.50 mCPU, slightly higher than Smart HPA's 1991.96 mCPU. However, the difference in CPU Overutilization remains significant, with Secure\-Smart HPA recording 156.00 mCPU compared to Smart HPA's 220.17 mCPU, achieving a 29.14\% improvement. While the Supply CPU is nearly the same, Smart HPA inefficiently allocates resources to microservices that do not need them. This inefficiency stems from the absence of disruption detection, and analysis, as well as the failure to adjust resource capacities accordingly.

\begin{tcolorbox}[left=0pt, top=0pt, right=0pt, bottom=0pt, boxrule=0.75pt]
Secure\-Smart HPA outperforms Smart HPA across all disruption levels by maintaining higher resource allocation and reducing CPU Overutilization. Through effective disruption monitoring, it adjusts microservice capacities, redistributes resources, and makes accurate auto-scaling decisions. In contrast, Smart HPA lacks disruption detection, leading to lower CPU supply and increased overutilization.
\end{tcolorbox}


\begin{figure*}[t]\vspace{-2mm}
    \begin{subfigure}[b]{0.32\textwidth} 
        \includegraphics[ trim=1cm 0.5cm 0.5cm 1cm, clip,width=1.1\textwidth]{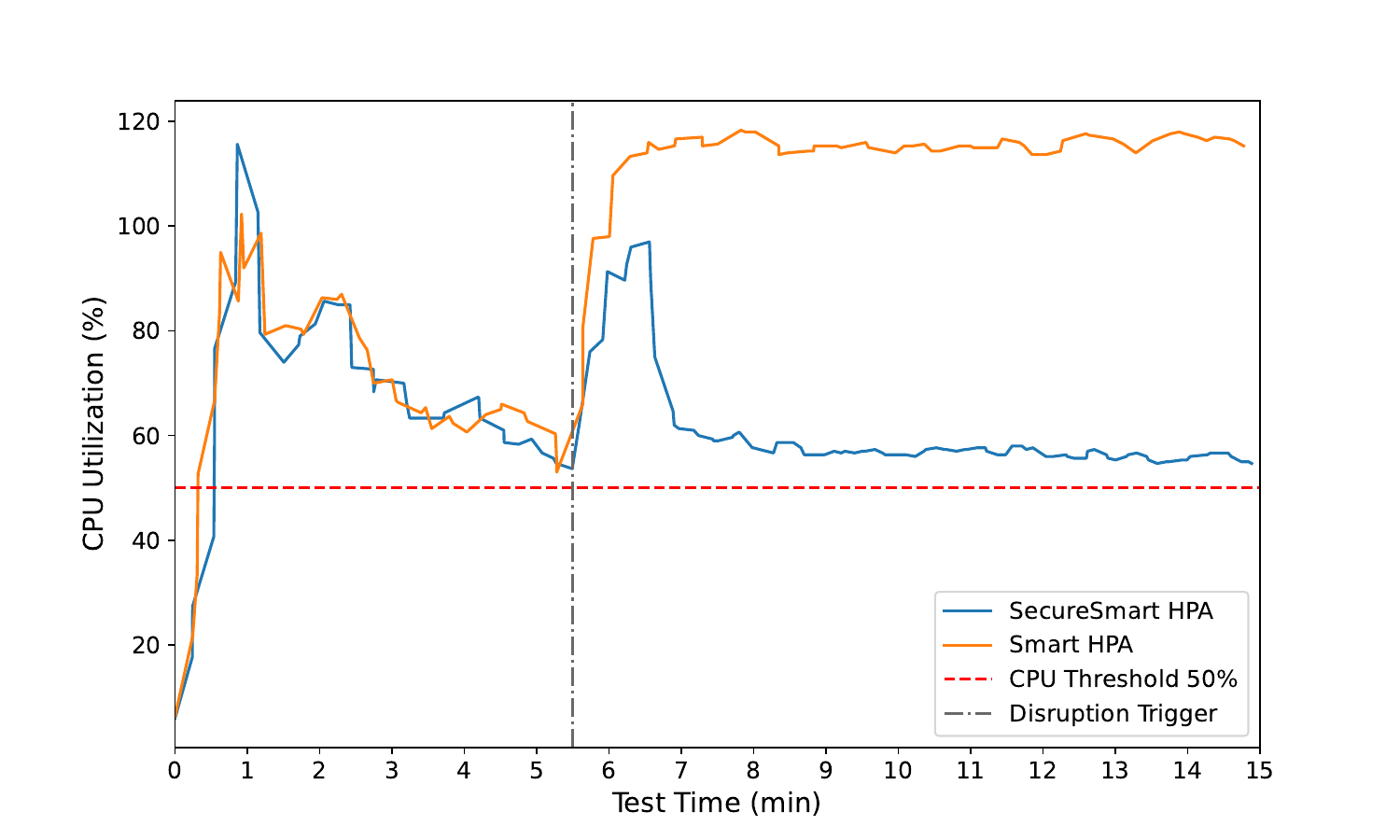} 
        \caption{Low Disruption Severity}
        \label{fig:fig1}
    \end{subfigure}
    \begin{subfigure}[b]{0.32\textwidth}
        \includegraphics[trim=1cm 0.5cm 0.5cm 1cm, clip,width=1.1\textwidth]{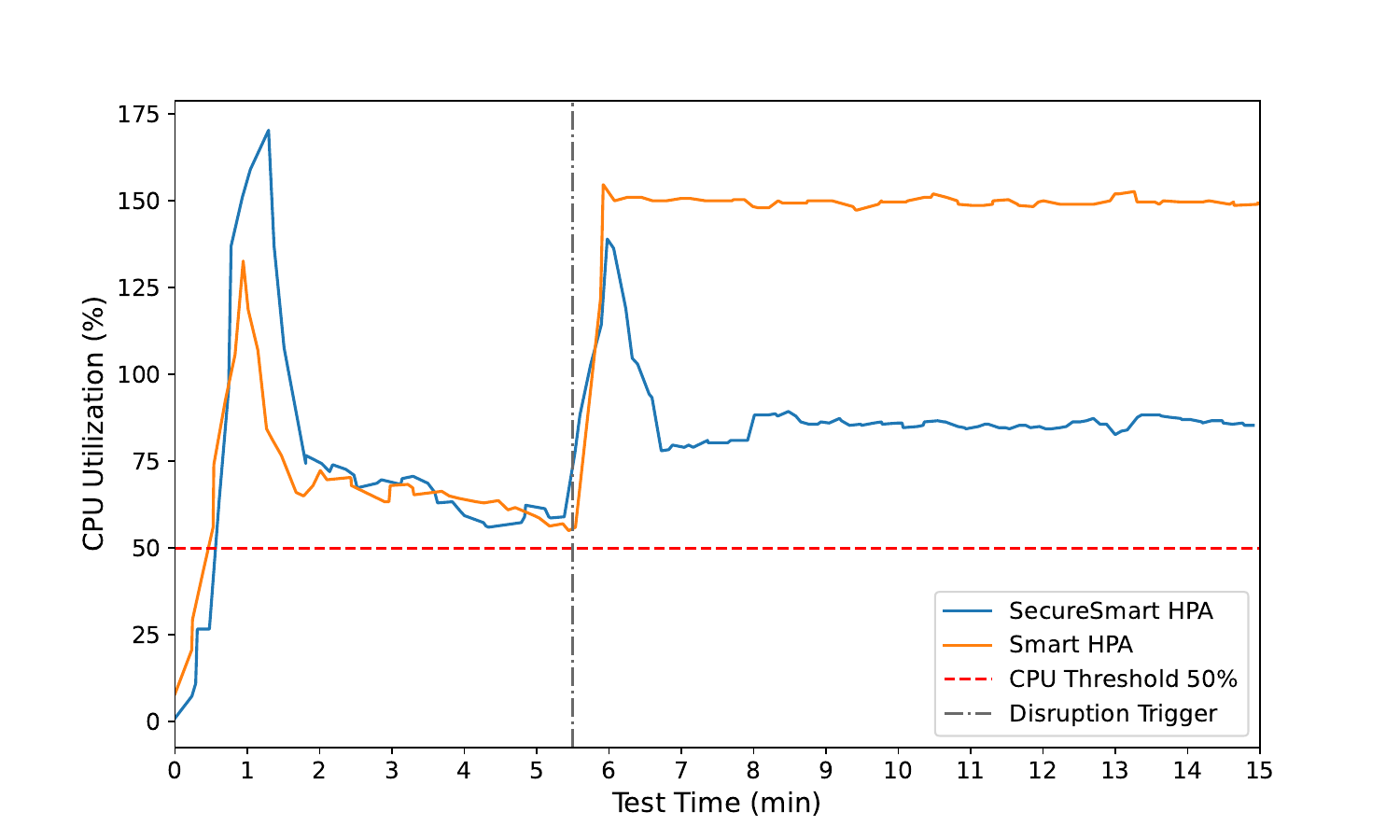} 
        \caption{Medium Disruption Severity}
        \label{fig:fig2}
    \end{subfigure}
    \begin{subfigure}[b]{0.32\textwidth}
        \includegraphics[trim=1cm 0.5cm 0.5cm 1cm, clip,width=1.1\textwidth]{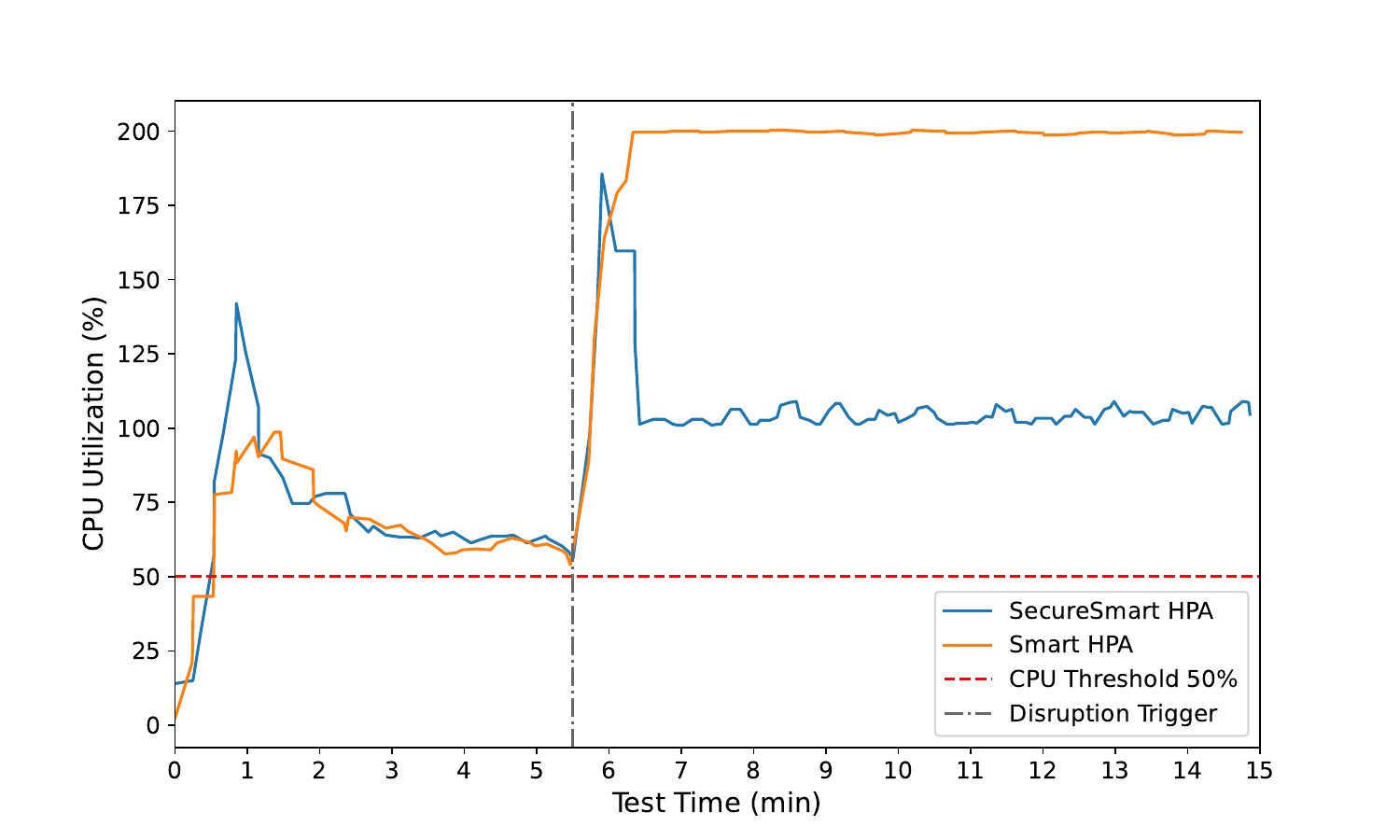} 
        \caption{High Disruption Severity}
        \label{fig:fig3}
    \end{subfigure}
    
    \caption{Secure\-Smart HPA vs. Smart HPA: CPU Utilization Comparison during Load Test.}
    \label{fig:results4}\vspace{1mm}%
\end{figure*}

\noindent \textbf{Secure\-Smart HPA vs. Smart HPA: Adaptive Behaviour Comparison.} Fig. \ref{fig:results4} shows the CPU utilization patterns of Secure\-Smart HPA and Smart HPA during a 15-minute load test across Low, Medium, and High disruption severity levels. Disruptions are introduced at the 5.5-minute mark by deleting microservice pods to simulate resource wastage for each severity level (Table~\ref{table:experimental_scenarios}). In the \textbf{Low severity} scenario (Fig.~\ref{fig:fig1}), both Secure\-Smart HPA and Smart HPA experience a spike in CPU utilization following the disruption. Secure\-Smart HPA stabilizes utilization around 60\%, remaining close to the 50\% threshold for the remainder of the test. In contrast, Smart HPA fluctuates between 110\% and 120\%, indicating persistent overutilization and inefficiency in resource management. Under \textbf{Medium severity} (Fig.~\ref{fig:fig2}), Secure\-Smart HPA exhibits a sharp spike in CPU utilization after the disruption but stabilizes around 90\%, maintaining a manageable level despite increased resource wastage. Smart HPA, however, operates at over 150\% CPU utilization throughout the test, reflecting severe overutilization and a lack of adaptability to resource constraints. In the \textbf{High severity} scenario (Fig.~\ref{fig:fig3}), Secure\-Smart HPA stabilizes CPU utilization at around 110\%, demonstrating resilience despite extreme resource loss. In contrast, Smart HPA operates at 200\% CPU utilization, indicating severe overutilization and ineffective disruption management.




\begin{tcolorbox}[left=0pt, top=0pt, right=0pt, bottom=0pt, boxrule=0.75pt]
Secure\-Smart HPA adapts to resource wastage caused by disruptions by striving to stabilize CPU utilization near the threshold level, maintaining it as close as possible. In contrast, Smart HPA, unable to account for resource wastage, fails to manage the disruption effectively, leading to increasing CPU utilization that drifts further from the threshold as the severity of disruption rises.
\end{tcolorbox}

\section{Threats to Validity} \label{section5}

We acknowledge several threats that may affect the generalizability of our findings. One limitation is the use of a single microservice benchmark application for evaluation, which could restrict the applicability of our results to other microservice architectures. However, given the widespread use, complexity, and technological diversity of the benchmark application, we believe our findings can extend to other microservice applications. Another threat is the reliance on default resource configurations of microservices of the benchmark application, such as resource requests and limits, which play a significant role in the observed improvements. Variations in these initial configurations may result in smaller or larger performance gains of Secure\-Smart HPA. Additionally, we compare Secure\-Smart HPA only with Smart HPA, rather than a broader set of alternative HPAs. Nonetheless, their structural and functional similarities, such as hierarchical architecture, threshold-based scaling policies, and resource exchange mechanisms, provide a strong basis for this comparison. Moreover, since Smart HPA outperforms Kubernetes baseline HPA \cite{ahmad2024smart}, its use as a benchmark reinforces the validity of Secure\-Smart HPA improvements. In addition, the performance of Secure\-Smart HPA may vary under different workload profiles. The selected profile, characterized by increasing and sustained high demand, reflects realistic traffic patterns and effectively simulates resource constraints, highlighting its potential for reliable performance in real-world scenarios. While the simulated resource disruptions align with real-world scenarios of resource wastage, their timing during load testing may influence the results, as earlier or later disruptions could affect the measured improvements over Smart HPA.




\section{Conclusion and Future Work} \label{section6}

This study introduces Secure\-Smart HPA, an adaptive, resilient, and resource-efficient HPA for microservice architectures. Built on a three-layered hierarchical MAPE-K architecture, it seamlessly integrates centralized and decentralized control frameworks while leveraging advanced auto-scaling heuristics to enhance the resilience and efficiency of microservice auto-scaling operations. Secure\-Smart HPA monitors resource demands, detects disruptions, assesses resource wastage, and dynamically adjusts scaling decisions, ensuring enhanced resilience and stability in dynamic microservice environments. Moreover, Secure\-Smart HPA maximizes resource efficiency by redistributing excess resources from overprovisioned to underprovisioned microservices in resource-constrained environments. Our experimental analysis demonstrates that Secure\-Smart HPA outperforms Smart HPA across various disruption severity levels, achieving up to a 57.2\% reduction in CPU overutilization and a 51.1\% increase in CPU allocation. 

Future improvements to Secure\-Smart HPA could focus on integrating predictive capabilities into Secure\-Smart HPA, such as leveraging AI or ML models to forecast disruptions and resource wastage. This would enable Secure\-Smart HPA to proactively adjust scaling decisions, enhancing its adaptability and efficiency in dynamic environments. Additionally, further evaluation of Secure\-Smart HPA could incorporate metrics such as stabilization time, response time, and cost-effectiveness to provide a more holistic view of Secure\-Smart HPA performance. Furthermore, Secure\-Smart HPA could be evaluated against diverse workload profiles and alternative scaling policies, such as queuing theory and fuzzy logic, to assess its flexibility and performance across different disruption scenarios.




\bibliographystyle{ieeetr}
\bibliography{ref}
\end{document}